\documentclass[%
reprint,
superscriptaddress,
amsmath,amssymb,
aps,
pra,
]{revtex4-2}
\usepackage{graphicx}
\usepackage{subfigure}
\usepackage{amsmath}
\usepackage{amssymb}
\usepackage{dcolumn}
\usepackage{bm}
\usepackage{blindtext}
\usepackage{listings}
\usepackage{color}
\usepackage{physics}
\usepackage[colorlinks,linkcolor=blue,anchorcolor=blue,citecolor=blue]{hyperref}
\usepackage{enumerate}
\makeatletter
\newif\if@restonecol
\makeatother

\usepackage[ruled,linesnumbered,vlined]{algorithm2e}
\usepackage{algpseudocode}

\DeclareMathOperator{\newE}{E}
\begin{document}
	\title{Accelerating Fermionic System Simulation on Quantum Computers }

\author{Qing-Song Li}
	\affiliation{Laboratory of Quantum Information, University of Science and Technology of China, Hefei 230026, China}
	\affiliation{Anhui Province Key Laboratory of Quantum Network, University of Science and Technology of China, Hefei 230026, China}

   \author{Jiaxuan Zhang}
    	\affiliation{Laboratory of Quantum Information, University of Science and Technology of China, Hefei 230026, China}
	\affiliation{Anhui Province Key Laboratory of Quantum Network, University of Science and Technology of China, Hefei 230026, China}
    
    \author{Huan-Yu Liu}
	\affiliation{Laboratory of Quantum Information, University of Science and Technology of China, Hefei 230026, China}
	\affiliation{Anhui Province Key Laboratory of Quantum Network, University of Science and Technology of China, Hefei 230026, China}
	
	\author{Qingchun Wang}
	\email{qingchun720@ustc.edu.cn}
	\affiliation{Institute of Artificial Intelligence, Hefei Comprehensive National Science Center, Hefei, Anhui 230026, China}
	
	\author{Yu-Chun Wu}
	\email{wuyuchun@ustc.edu.cn}
	\affiliation{Laboratory of Quantum Information, University of Science and Technology of China, Hefei 230026, China}
	\affiliation{Anhui Province Key Laboratory of Quantum Network, University of Science and Technology of China, Hefei 230026, China}
	
	\affiliation{Institute of Artificial Intelligence, Hefei Comprehensive National Science Center, Hefei, Anhui 230026, China}
	
	\author{Guo-Ping Guo}
	\affiliation{Laboratory of Quantum Information, University of Science and Technology of China, Hefei 230026, China}
	\affiliation{Anhui Province Key Laboratory of Quantum Network, University of Science and Technology of China, Hefei 230026, China}
	\affiliation{Institute of Artificial Intelligence, Hefei Comprehensive National Science Center, Hefei, Anhui 230026, China}
	\affiliation{Origin Quantum Computing, Hefei, Anhui 230026, China}

	\date{\today}
	\begin{abstract}
A potential approach for demonstrating quantum advantage is using quantum computers to simulate fermionic systems. Quantum algorithms for fermionic system simulation usually involve the Hamiltonian evolution and measurements. However, in the second quantization representation, the number of terms in many fermion-system Hamiltonians, such as molecular Hamiltonians, is substantial, approximately $\mathcal{O}(N^4)$, where $N$ is the number of molecular orbitals. Due to this, the computational resources required for Hamiltonian evolution and expectation value measurements could be excessively large.
To address this, we introduce a grouping strategy that partitions these $\mathcal{O}(N^4)$ Hamiltonian terms into $\mathcal{O}(N^2)$ groups, with the terms in each group mutually commuting. Based on this grouping method, we propose a parallel Hamiltonian evolution scheme that reduces the circuit depth of Hamiltonian evolution by a factor of $N$. Moreover, our grouping measurement strategy reduces the number of measurements needed to $\mathcal{O}(N^2)$, whereas the current best grouping measurement schemes require $\mathcal{O}(N^3)$ measurements. Additionally, we find that measuring the expectation value of a group of Hamiltonian terms requires fewer repetitions than measuring a single term individually, thereby reducing the number of quantum circuit executions.
 Our approach saves a factor of $N^3$ in the overall time for Hamiltonian evolution and measurements, significantly decreasing the time required for quantum computers to simulate fermionic systems.
	\end{abstract}
	
	\maketitle
	\section{Introduction}
       In the 1980s, Feynman proposed using quantum computers to simulate quantum systems~\cite{feynmansimulating}. One prominent example is the quantum phase estimation (QPE) algorithm~\cite{babbush2018low,reiher2017elucidating,cao2018potential,kassal2011simulating,lu2012quantum,aspuru2018matter}, which can be used to calculate the ground state energy of a quantum system. Despite technological advancements that have led to increased qubit counts and higher operational fidelity~\cite{fowler2012surface}, implementing the QPE algorithm remains highly challenging due to its substantial circuit depth. To more rapidly demonstrate the advantages of quantum computing, a quantum variational algorithm, the variational quantum eigensolver (VQE)~\cite{peruzzo2014variational,mcclean2016theory,2023Noisy,2023Modular}, has been proposed. This algorithm offers the benefit of requiring shorter circuits and, due to its variational nature, exhibits greater tolerance to noise, making it more suitable for current quantum computing hardware. 
       
       Hamiltonian simulation and expectation value measurements are subroutines in many quantum algorithms for quantum chemistry, such as QPE and VQE. Nonetheless, in the second quantization representation, the number of terms in a molecular Hamiltonian is $\mathcal{O}(N^4)$. Using Trotter decomposition~\cite{trotter1959product,suzuki1992general} to evolve the molecular Hamiltonian results in a circuit depth of $\mathcal{O}(N^4 \log(N) M)$, where $\log(N)$ represents the depth required to evolve a single Hamiltonian term, and $M$ is the number of Trotter steps. The number of Trotter steps depends on the required evolution precision and the Hamiltonian norm. Such deep quantum circuits pose significant challenges for execution on current quantum computing hardware. After the Hamiltonian evolution, the next step is typically to measure its expectation value. If each term is measured individually, the total number of measurements required scales as $\mathcal{O}(N^4 N_m)$, where $N_m$ is the number of measurements required for each term, which is determined by the desired precision and the variance of that term. 

The total computational time is given by the product of the time taken to execute a single circuit, $\mathcal{O}(N^4 \log(N) M)$, and the number of circuits, $\mathcal{O}(N^4 N_m)$. This results in an overall time complexity of $\mathcal{O}(N^8\log(N)MN_m)$. As is evident, this total computational cost is prohibitively high. Therefore, it is essential to reduce both the circuit depth required for Hamiltonian evolution and the number of measurements needed to evaluate the Hamiltonian expectation value.
        
        Currently, a widely used method to reduce measurement resources is the classical shadows technique~\cite{huang2019predictingfeaturesquantumsystems,Hadfield2022,Elben2023, Gresch2025}. The number of measurements required is $\mathcal{O}(\frac{\log(M)}{\epsilon^2}\text{max}_i3^{w_i})$, where $M$ is the number of terms and $\epsilon$ is the measurement error for each term. Here, $w_i$ represents the Pauli weight of the $i$-th term. This type of scheme is effective for local observables with \(w = \mathcal{O}(1)\). However, for fermionic systems, if we apply the Jordan-Wigner transformation (JWT)~\cite{jordan1928pauli} to map the fermionic Hamiltonians to Pauli Hamiltonians, the maximum Pauli weight of Pauli Hamiltonian terms is $N$. Consequently, the number of measurements increases exponentially with $N$. If the optimal ternary-tree transformation~\cite{jiang2020optimal} is used instead, the maximum $w_i$ of Pauli Hamiltonian terms is reduced to $4 \log_3(N)$. In this case, the number of measurements is also $\mathcal{O}(N^4)$ in terms of $N$, which has the same complexity as the individual measurement.

        Some works~\cite{huggins2021efficient,Motta2021,Peng2017} employ algebraic decomposition methods to transform the fermionic Hamiltonian. By truncating the expansion terms, the number of Hamiltonian terms can be reduced to $\order{N}$ at most, thereby reducing the number of measurements. Since the decomposed operators are diagonal, the Hamiltonian simulation circuits can be significantly simplified. However, the truncation error is difficult to estimate, and when the system size is large, truncating to $\order{N}$ terms may not be sufficient.
        
        Another class of strategies for reducing measurement resources is grouping measurement based on greedy algorithms~\cite{romero2018strategies,hamamura2020efficient,yen2020measuring,choi2022improving,gokhale2020n,javadi2024quantum,jena2022optimization}. This approach begins by constructing a graph where vertices represent the $\mathcal{O}(N^4)$ Hamiltonian terms. An edge is drawn between two vertices if the two corresponding terms commute. A greedy algorithm is then employed to partition the graph into fully-connected subgraphs, allowing Hamiltonian terms within these subgraphs to be measured simultaneously. However, this method incurs high complexity, as the construction of the graph alone requires $\mathcal{O}(N^8)$ operations~\cite{verteletskyi2020measurement}. More critically, the fact that these greedy algorithms~\cite{leighton1979graph,dutton1981new,hertz1990fast} usually require a quadratic or cubic runtime with respect to the number of vertices implies a scaling of $\mathcal{O}(N^8)$ or $\mathcal{O}(N^{12})$ in classical precomputation time. For molecules with hundreds of orbitals, these algorithms become impractical for classical computers to execute within a reasonable timeframe. The optimal number of groups generated by grouping strategies based on greedy algorithms is estimated to scale as $\mathcal{O}(N^3)$~\cite{yen2020measuring,gokhale2020n,jena2022optimization}. There are also some workss~\cite{PhysRevA.101.062322, Shlosberg2023adaptiveestimation,Wu2023overlappedgrouping,Crawford2021efficientquantum} dedicated to reducing the number of measurements, $N_m$, for each operator or each group of operators.

      We adopt a Hamiltonian grouping approach that simultaneously reduces both the circuit depth for Hamiltonian evolution and the number of measurements required. By applying fermion-to-qubit mappings to the molecular Hamiltonian, we obtain a Pauli Hamiltonian. Leveraging its structural properties, we propose an efficient grouping scheme that does not require constructing a commutativity graph and instead directly partitions the $\mathcal{O}(N^4)$ Hamiltonian terms into $\mathcal{O}(N^2)$ groups based on specific rules. As a result, the classical computational resources required for this grouping are $\mathcal{O}(N^4)$, significantly lower than those required by other grouping schemes. Furthermore, the number of groups can be deterministically reduced to $\mathcal{O}(N^2)$.

Based on this grouping scheme, we propose a parallel Hamiltonian evolution method that reduces the circuit depth of Trotter-decomposition-based Hamiltonian evolution to $\mathcal{O}(N^3 \log (N) M)$. By simultaneously measuring the expectation value of each group, the total number of required measurements is reduced to $\mathcal{O}(N^2 N_m^\prime)$. Moreover, our numerical tests indicate that the number of quantum circuit executions (shots) per group,  $N_m^\prime$, is significantly lower than that required when measuring Hamiltonian terms individually.

        The rest of this paper is organized as follows:
        In Sec.~\ref{background}, we introduce the second quantization and the Jordan-Wigner transformation, followed by a discussion on Hamiltonian evolution and expectation value measurements. In Sec.~\ref{gph}, we provide a detailed description of the Hamiltonian grouping scheme. In Sec.~\ref{psh}, the parallel Hamiltonian simulation scheme is presented. In Sec.~\ref{smg}, we describe the method for simultaneously measuring groups of Hamiltonian terms. In Sec.~\ref{ntr}, we will present the numerical test results. Finally, a summary is provided in Sec.~\ref{cad}.

	\section{Background}\label{background}
	\subsection{Second quantization}\label{SecondQuantization}

In the second quantization representation, the Fock space of $N$ orbitals is spanned by the following set of basis states
        \begin{equation}
            \{\ket{n_1, n_2, \cdots, n_j, \cdots, n_{N}} \},
        \end{equation}
where $n_j \in \{0, 1\}$ denotes the occupation number of the $j$-th orbital. Due to the Pauli exclusion principle, each orbital can accommodate at most one fermion.

Operators in fermionic systems can be expressed in terms of the fundamental creation and annihilation operators, defined as
        \begin{equation}
            \begin{split}
             \begin{split}\label{defing_fer_op}
			a_j\ket{n_1,\dots,n_j,\dots,n_{N}}=\Gamma_j \delta_{1,n_j}\ket{n_1,\dots,0,\dots,n_{N}},\\
			a_j^\dagger \ket{n_1,\dots,n_j,\dots,n_{N}}=\Gamma_j \delta_{0,n_j}\ket{n_1,\dots,1,\dots,n_{N}} ,
		\end{split}   
            \end{split}
        \end{equation}
        where $\Gamma_j = (-1)^{\sum_{k=1}^{j-1}\hat{n}_k}$ is the phase factor, and $\hat{n}_j=a_j^\dagger a_j$ is the particle number operator. The creation and annihilation operators satisfy the following relations
        \begin{equation}
		\begin{split}
			&\{a_i,a_j\}=0,\\
			&\{a_i^\dagger,a_j^\dagger\}=0,\\
			&\{a_i,a_j^\dagger\}=\delta_{i,j},
		\end{split}
  \end{equation}
where $\{A, B\} \equiv AB + BA$.

    The molecular electronic structure Hamiltonian can be expressed in terms of creation and annihilation operators as 
    \begin{equation}
        H = \sum_{p,q} h_{ij}a_i^\dagger a_j + \frac{1}{2}\sum_{p,q,s,r}h_{pqsr}a_p^\dagger a_q^\dagger a_r a_s,
    \end{equation}
where $h_{ij}$ and $h_{pqrs}$ are the one-electron and two-electron integrals, respectively. Many chemical software packages ~\cite{wang2021chemiq,cross2018ibm,mcclean2020openfermion} can compute the molecular electronic structure Hamiltonian. In this work, we use the PyChemiQ package~\cite{wang2021chemiq} to obtain the Hamiltonian.

    \subsection{Jordan-Wigner transformation}
  Fermionic operators cannot be directly executed on a quantum computer, so fermion-to-qubit mappings are needed. Many fermion-to-qubit mappings have been proposed~\cite{bravyi2002fermionic,seeley2012bravyi,tranter2015b,somma2002simulating,nielsen2005fermionic,whitfield2011simulation,li2022unified,miller2023bonsai,chen2024error}, and the earliest and most widely used mapping is the Jordan-Wigner transformation (JWT). In this work, we utilize the JWT primarily and will provide an overview of this transformation. It is important to note that using different mappings does not affect our grouping results, as the commutation relations between operators remain independent of the basis.
  
    The JWT employs $N$ qubits to record the occupation numbers of the corresponding orbitals
    \begin{equation}
          \ket{n_1,\cdots, n_j, \cdots, n_{N}} \rightarrow \ket{q_1, \cdots, q_j, \cdots, n_{N}}, 
    \end{equation}
    where $q_j = n_j$. Creation and annihilation  operators in the JWT can be expressed as 
    \begin{equation}
        \begin{split}
            &a_j \rightarrow Z_1 Z_2 \cdots Z_{j-1} \frac{X_j + iY_j}{2},\\
            &a_j^\dagger \rightarrow Z_1 Z_2 \cdots Z_{j-1} \frac{X_j - iY_j}{2}.
        \end{split}
    \end{equation}

    Using the JWT, a fermionic Hamiltonian can be expressed as a Pauli Hamiltonian, which is a linear combination of multi-qubit Pauli operators
    \begin{equation}
        H_{\text{P}} = \sum_l  h_lH_l, H_l = \prod_j P_j, 
    \end{equation}
    where $P_j \in \{X_j, Y_j, Z_j\}$. Note that the identity operator is omitted for simplicity.

    \subsection{Hamiltonian simulation circuits}\label{hsc}
    In many quantum algorithms, converting the operator $e^{-i h P}$ into a quantum circuit is a common task, where $h$ is a coefficient and $P$ is a multi-qubit Pauli operator. To perform this conversion, we first transform the Pauli $X$ and $Y$ operators in $P$ into Pauli $Z$ operators using appropriate rotation gates. Using the following identities
    \begin{equation}
        HXH = Z, \quad R_XYR_X^\dagger = Z,
    \end{equation}
   we apply the Hadamard gates to the qubits where Pauli $X$ operators act and apply the $R_X$ gates to the qubits where Pauli $Y$ operators act. The $R_X$ gate is a single-qubit rotation gate around the X-axis, and the rotation angle is $\pi/2$. Next, we accumulate the parity of all qubits on which $P$ acts, which can be carried out using $CNOT$ gates. For example, $CNOT(i, j)$ can load the parity of the $i$-th qubit to the $j$-th qubit. Once the parity has been consolidated onto one qubit, we apply the $ R_Z(\theta) $ gate on that qubit, where $\theta = \frac{h}{2}$ represents the rotation angle around the Z-axis. After performing the rotation, we reverse the parity accumulation process using $CNOT$ gates again. Finally, we apply basis change operations on the affected qubits to recover their original directions. 
\begin{figure}[h]
		\centering

		\subfigure[]{
			\includegraphics[width=0.4\textwidth]{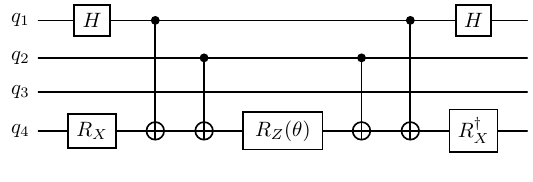}\label{qc_sim}    }
		\subfigure[]{
			\includegraphics[width=0.45\textwidth]{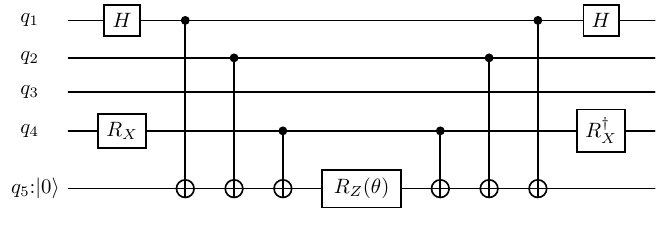} \label{qc_sim_aux}   }
       \subfigure[]{
    			\includegraphics[width=0.225\textwidth]{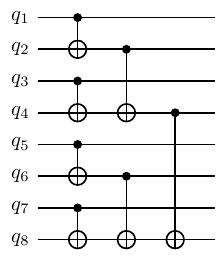}\label{qc_load_pa}    }
                  \subfigure[]{
                	\includegraphics[width=0.2\textwidth]{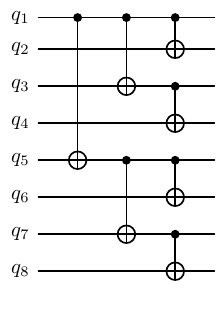}\label{qc_copy_pa}    }
		\caption{ The circuits to simulate the evolution of the one-term Hamiltonian $e^{-ihP}$, where $P= X_1 Z_2 Y_4$. a) A circuit for simulating a single Hamiltonian term. b) A circuit for simulating the single Hamiltonian term with an auxiliary qubit. c) A circuit designed for accumulating parity in parallel. d) A circuit developed for parallel copying of parity. }
	\end{figure}
	
For instance, if $P$ is $X_1 Z_2 Y_4$, the circuit implementing this operation is illustrated in Figure~\ref{qc_sim}. Notably, the parity can be accumulated onto an auxiliary qubit initialized to $\ket{0}$, and performing the rotation on this qubit can also realize $e^{-i h P}$, as shown in Figure~\ref{qc_sim_aux}. The number of gates required to implement the evolution circuit is $\mathcal{O}(wt(P))$, where $wt(P)$ denotes the Pauli weight of $P$, defined as the total number of Pauli $X$, $Y$, and $Z$ operators in $P$. The depth of the circuit in this approach is also $\mathcal{O}(wt(P))$. However, by parallelizing the parity accumulation, we can effectively reduce the circuit depth. Figure~\ref{qc_load_pa} shows a circuit for parallel parity accumulation for 8 qubits. This approach reduces the circuit depth to $\log(wt(P))$. Similarly, the parity of one qubit can also be copied to other qubits in parallel, as shown in Figure~\ref{qc_copy_pa}.
    
 \subsection{Hamiltonian expectation value measurements}
In many quantum algorithms for solving fermionic systems, measuring the expectation value of the Hamiltonian $H_P$ is a crucial step. The direct approach is measuring the expectation value of each term in the Hamiltonian separately and then summing these expectation values to obtain the energy of the Hamiltonian
    \begin{equation}
        \newE{(H_P)} = \newE{(\sum_l h_l H_l)} = \sum_l h_l\newE{(H_l)}.
    \end{equation}
   
 If a Pauli $Z$ operator acts on the $j$-th qubit, a Z-direction measurement is performed on that qubit. If a Pauli $X$ or $Y$ operator acts on the $j$-th qubit, a corresponding Hadamard gate or $R_X$ gate is applied, followed by a measurement in the Z-direction on that qubit.

To reduce the number of measurements required for estimating the Hamiltonian expectation value, mutually commuting Hamiltonian terms can be measured simultaneously. Operators that commute with each other are referred to as fully-commuting operators, such as $X_1X_2$ and $Y_1Y_2$. A stricter type of commutativity, called qubit-wise commuting (QWC), requires that each pair of operators commute on every qubit. For example, $X_1Z_2$ and $Z_2Y_3$ are qubit-wise commuting, whereas $X_1X_2$ and $Y_1Y_2$ are not qubit-wise commuting.

Qubit-wise commuting operators can be measured simultaneously using the using the method described earlier. In contrast, fully-commuting operators require an auxiliary circuit to transform them into qubit-wise commuting operators before measurement. The construction of the auxiliary circuit will be described in Sec.~\ref{psh}.

    \section{Grouping Pauli Hamiltonians}\label{gph}
    The fermionic Hamiltonian consists of two-body terms, $a_i^\dagger a_j$, and four-body terms, $a_p^\dagger a_q^\dagger a_s a_r$. We now explicitly compute their corresponding multi-qubit Pauli operators under the Jordan-Wigner transformation. For convenience, a multi-qubit Pauli operator is often referred to as a Pauli string.
    
    First, consider the two-body term $a_i^\dagger a_j$ with $i < j$. Under the JWT, this term results in four Pauli strings: $X_i Z_{i+1} \cdots Z_{j-1} X_j$, $X_i Z_{i+1} \cdots Z_{j-1} Y_j$, $Y_i Z_{i+1} \cdots Z_{j-1} X_j$, and $Y_i Z_{i+1} \cdots Z_{j-1} Y_j$. Since the fermionic Hamiltonian always contains the Hermitian conjugate term ($a_j^\dagger a_i$) of $a_i^\dagger a_j$, summing these terms results in the cancellation of $X_i Z_{i+1} \cdots Z_{j-1} Y_j$ and $Y_i Z_{i+1} \cdots Z_{j-1} X_j$, leaving only $X_i Z_{i+1} \cdots Z_{j-1} X_j$ and $Y_i Z_{i+1} \cdots Z_{j-1} Y_j$. For simplicity, we refer to these as the $XX$-type and $YY$-type Pauli strings, respectively. If $i = j$, only two Pauli strings, $Z_j$ and $I$~(the identity operator), are generated. We denote these two types of Pauli strings as $Z$-type and $I$-type, respectively. In summary, the two-body terms either generate Pauli strings that begin and end with Pauli $X$ ($Y$) operators, with a continuous Pauli $Z$ operators between Pauli $X$ ($Y$) opertors, or they generate single-qubit Pauli $Z$ ($I$) operators.

   The four-body terms can be viewed as the product of two two-body terms. Consider the case where the four indices $p, q, r, s$ are all distinct. Without loss of generality, assume $p < q < r < s$. Then, the term $a_p^\dagger a_q^\dagger a_s a_r$ will be transformed into the summation of 16 Pauli strings similar to the form $X_p Z_{p+1} \cdots Z_{q-1} X_q X_r Z_{r+1} \cdots Z_{s-1} X_s$. We categorize these types of Pauli strings as the $XXXX$-type. Note that $X$ can be replaced by $Y$, resulting in a total of 16 types, such as $XYXY$- and $XYYX$-types, etc. However, due to the symmetry of the Hamiltonian, many of these types cancel out, leaving the following types: $XXXX$-, $YYYY$-, $XXYY$-, $YYXX$-, $XYYX$-, and $YXXY$-types.

Next, consider the case where $p = q < r < s$. The corresponding terms under the JWT will generate Pauli strings similar to $Z_p X_r Z_{r+1} \cdots Z_{s-1} X_s$, denoted as the $ZXX$-type, as well as $X_r Z_{r+1} \cdots Z_{s-1} X_s$, which is already generated in the two-body terms. By replacing $X$ with $Y$ in the $ZXX$-type, we also obtain the $ZYY$-type Pauli strings. Due to symmetry, the $ZXY$- and $ZYX$-types Pauli strings will be eliminated. Similarly, if $p < q < r = s$, two new types of Pauli strings generated are the $XXZ$- and $YYZ$-types. 

If the two equal indices occur in the middle: $p < q = r < s$, new operators will be generated that resemble $X_p Z_{p+1} \cdots Z_{q-1} Z_{q+1} \cdots X_s$. The newly generated types of operators are denoted as $XZX$- and $YZY$-types, where $Z$ represents a gap appearing in the middle of a continuous string of Pauli $Z$ operators. For cases with three identical indices or four identical indices, no new types of Pauli strings are generated.

To summarize the results above, the Pauli strings obtained from the fermionic Hamiltonian under the JWT can be categorized into 16 types, as shown in Table~\ref{tableofPauliClass}. 
    
    \begin{table}[!htbp]
		\centering
		\caption{The types of Pauli strings under the Jordan-Wigner transformation.\label{tableofPauliClass}}
		\begin{tabular}[c]{cc}
			\hline \hline
			Pauli Type			&Pauli String	\\
			\hline
			$I$&       $I$		\\
                $Z$&       $Z_i$    \\
                $ZZ$&       $Z_iZ_j$    \\
                $XX$&       $X_iZ_{i+1}\cdots Z_{j-1}X_j$   \\
                $YY$&       $Y_iZ_{i+1}\cdots Z_{j-1}Y_j$   \\
                $ZXX$&      $Z_i X_jZ_{j+1}\cdots Z_{k-1}X_k$   \\
                $ZYY$&      $Z_i Y_jZ_{j+1}\cdots Z_{k-1}Y_k$   \\
                 $XXZ$&      $X_iZ_{i+1}\cdots Z_{j-1}X_jZ_k$   \\
                $YYZ$&      $Y_iZ_{i+1}\cdots Z_{j-1}Y_jZ_k$   \\
                $XZX$&      $X_iZ_{i+1}\cdots Z_{j-1}Z_{j+1}\cdots  Z_{k-1}X_k$   \\
                $YZY$&      $Y_iZ_{i+1}\cdots Z_{j-1}Z_{j+1}\cdots  Z_{k-1}Y_k$   \\
                $XXXX$&     $X_iZ_{i+1}\cdots Z_{j-1}X_j X_kZ_{k+1}\cdots Z_{l-1}X_l$\\
                $YYYY$&     $Y_iZ_{i+1}\cdots Z_{j-1}Y_j Y_kZ_{k+1}\cdots Z_{l-1}Y_l$\\
                $XXYY$&     $X_iZ_{i+1}\cdots Z_{j-1}X_j Y_kZ_{k+1}\cdots Z_{l-1}Y_l$\\
                $YYXX$&     $Y_iZ_{i+1}\cdots Z_{j-1}Y_j X_kZ_{k+1}\cdots Z_{l-1}X_l$\\
                $XYYX$&     $X_iZ_{i+1}\cdots Z_{j-1}Y_j Y_kZ_{k+1}\cdots Z_{l-1}X_l$\\
                $YXXY$&     $Y_iZ_{i+1}\cdots Z_{j-1}X_j X_kZ_{k+1}\cdots Z_{l-1}Y_l$\\
			\hline \hline
		\end{tabular}
	\end{table}

 For simplicity, in the following content, consecutive $Z$ operators will be omitted when writing the multi-qubit Pauli operators in Table~\ref{tableofPauliClass}. For example, $Z_i Y_j Z_{j+1} \cdots Z_{k-1} Y_k$ will be abbreviated as $\widehat{Z_i Y_j Y_k}$, and $Y_i Z_{i+1} \cdots Z_{j-1} X_j X_k Z_{k+1} \cdots Z_{l-1} Y_l$ will be written as $\widehat{Y_i X_j X_k Y_l}$. Specifically, $X_i Z_{i+1} \cdots Z_{j-1} Z_{j+1} \cdots Z_{k-1} X_k$ can be viewed as $X_i Z_j X_k \times Z_{i+1} \cdots Z_{j-1}Z_j Z_{j+1}\cdots Z_{k-1}$. Similarly, we omit the consecutive $Z$ operators, and thus $X_i Z_{i+1} \cdots Z_{j-1} Z_{j+1} \cdots Z_{k-1} X_k$ is abbreviated as $\widehat{ X_i Z_j X_k }$.

    Now, we group these Pauli strings to ensure that all Pauli strings within each group commute with each other. In this way, expectation values of the Pauli strings within a group can be measured simultaneously. The Pauli strings of  $I$-, $Z$-, and $ZZ$-types all commute with each other. We can group them together, and denote this group as
    \begin{equation}
         G_1 = \{I, Z_i, Z_iZ_j| 1 \leq i < j \leq N\}
    \end{equation}
    
   Then we consider the $ZXX$-, $ZYY$-, $XZX$-, $YZY$-, $XXZ$-, and $YYZ$-types. It can be observed that the Pauli strings of these types each contain two $X$ or $Y$ operators. We group the Pauli strings in which the indices of the first $X$ $(Y)$ operator and the second $X$ $(Y)$ operator are the same. We denote this group as
   \begin{equation}
   \begin{split}
         G_2(a,b) =\{&\widehat{Z_iX_aX_b},\widehat{X_aZ_jX_b},\widehat{X_aX_bZ_k}, \\
                &\widehat{Z_iY_aY_b},\widehat{Y_aZ_jY_b},\widehat{Y_aY_bZ_k}   \\
                &|1 \leq i < a <j < b < k \leq N \}.
   \end{split}
   \end{equation}
   For example, $Z_1 X_3 Z_4 X_5$ and $X_3 Z_4 X_5 Z_6$ belong to the group $G_2(3,5)$. 
   
   We define a Pauli string as an $n$-body term if there are $n$ non-$Z$ (including both $X$ and $Y$ operators) operators in this Pauli string. Then we consider how to group the four-body terms. First, consider the operators of the $XXXX-$ and $YYYY$-types. Suppose the indices of the Pauli $X$ or $Y$ operators in a Pauli string of these types are $i$, $j$, $k$, and $l$. Based on the four indices, we define two new variables: $Med$ and $Bia$. Their definitions are as follows
   \begin{equation}
       Med = \frac{j+k}{2},\quad Bia = (l-k) -(j-i).
   \end{equation}
   
    The value of $Med$ can be either an integer or a half-integer, and the number of possible values for $Med$ does not exceed $2N$. Similarly, the number of possible values for $Bia$ also does not exceed $2N$. 

With these two variables defined, the grouping of $XXXX$- and $YYYY$-types becomes straightforward. The Pauli strings of these types can be grouped based on their corresponding $Med$ and $Bia$ values,
    \begin{equation}
         \begin{split}
         & G_3(m,b) = \{\widehat{X_iX_jX_kX_l}|Med = m, Bia = b\},\\
          &G_4(m,b) = \{\widehat{Y_iY_jY_kY_l}|Med = m, Bia = b\}.
         \end{split}
    \end{equation}

The method for grouping $XXYY$- and $YYXX$-types is the same as that for $XXXX$- and $YYYY$-types
     \begin{equation}
     \begin{split}
        &G_5(m,b) = \{\widehat{X_iX_jY_kY_l}|Med = m, Bia = b\},\\
         &G_6(m,b) = \{\widehat{Y_iY_jX_kX_l}|Med = m, Bia = b\}.
     \end{split}
    \end{equation}

The method for grouping Pauli strings of $XYYX$- and $YXXY$-types differs from the previous methods. Two new variables need to be defined
    \begin{equation}
        Med_1 = \frac{i+j}{2},\quad Med_2 = \frac{k+l}{2}.
    \end{equation}
   Then, the Pauli strings of $XYYX$- and $YXXY$-types are grouped by the two new variables $Med_1$ and $Med_2$
    \begin{equation}
    \begin{split}
       & G_7(m_1,m_2) = \{\widehat{X_iY_jY_kX_l}|Med_1 = m_1, Med_2 = m_2\},\\
       & G_8(m_1,m_2) = \{\widehat{Y_iX_jX_kY_l}|Med_1 = m_1, Med_2 = m_2\}.
    \end{split}
    \end{equation}
    We will demonstrate in Appendix~\ref{dis_scheme} that the operators within each group commute with each other.
    
 We calculate the number of groups in a Hamiltonian carefully. Obviously, The number of $G_1$ is 1. Then we consider the ranges of the parameters $a$ and $b$ in $G_2(a,b)$. Parameters $a$ and $b$ are integers, with their values in the range $[1, N]$. Actually, $a$ cannot be equal to $N$ and $b$ cannot be equal to 1, but for convenience in calculation, we expand the range to $[1,N]$ and apply similar expansion in the subsequent calculations. Therefore, the number of $G_2$ is less than $N^2$. Next, consider the quantities of $G_3$ to $G_6$. $Med$ represents the symmetry axis, which can take integer and half-integer values within the range $[5/2, (2N-3)/2]$. Thus, the number of possible values for $Med$ is less than $2N$. The parameter $Bia$ represents the bias of the length of two substrings, which can only take integer values but may be negative. According to the definition, the range of $Bia$ is $[4-N, N-4]$ (assuming $N \leq 4$), and we also expand its value range to $[-N, N]$. Therefore, the total number of $G_3$, $G_4$, $G_5$, and $G_6$ is less than $4 \times (2N) \times (2N) = 16N^2$. Similar to $Med$, the number of possible values for $Med_1$ and $Med_2$ is also less than $2N$. Thus, the total number of $G_7$ and $G_8$ is less than $2 \times (2N) \times (2N) = 8N^2$. In total, the number of all groups is less than $25N^2 + 1$.

In summary of the grouping scheme introduced above, it is clear that the scaling of the total number of groups in a molecular Hamiltonian is $\mathcal{O}(N^2)$. Our grouping scheme utilizes certain properties of the JWT. Although other transformations do not share these same favorable properties, we can use the grouping results of the JWT to guide the grouping methods for other mappings. For example, if two terms, $P_\text{JW}$ and $P_\text{JW}^\prime$, belong to the same group, then the corresponding terms $P_\text{BK}$ and $P_\text{BK}^\prime$ under the Bravyi-Kitaev transformation (BKT) can also be categorized into one group, as different mappings do not alter the commutation relations of the two operators. Specifically, the BKT can be related to the JWT through a Unitary transformation
\begin{equation}
   \ket{\psi}_{BK} = U \ket{\psi}_{JW}, \quad H_{BK} = U H_{JW} U^\dagger.
\end{equation}
If \( P_{JW} \) and \( P'_{JW} \) commute: $[P_{JW}, P'_{JW}] = 0$ , then \( P_{BK} \) and \( P'_{BK} \) also commute, because
\begin{equation}
\begin{split}
    [P_{BK}, P'_{BK}] &= [U P_{JW} U^\dagger, U P'_{JW} U^\dagger]\\
    &= U [P_{JW}, P'_{JW}] U^\dagger\\
    &=0.
\end{split}
\end{equation}
When grouping the BK Hamiltonian, we first group the JW Hamiltonian, and then group the BK Hamiltonian according to the results of the JW Hamiltonian grouping. If a term $H_{l,JW}$ in the JW Hamiltonian is assigned to a group $G_{n,JW}$, the corresponding term $H_{l,BK} = UH_{l,JW}U^\dagger$ in the BK Hamiltonian is assigned to the correspongding group $G_{n,BK}$. Since this $U$  can be decomposed into the Clifford circuit, it is straightforward to compute $UH_{l,JW}U^\dagger$ on classical computers.
For example, we consider the 4-orbital  $\text{H}_2$ molecule. In this case, the $U$ can be decomposed as $CNOT(1,2) CNOT(2,4)CNOT(3,4)$. The terms, $Z_1$ and $Z_2$ , in the JW $\text{H}_2$ Hamiltonian are assigned to  group \( G_{1,JW} \), then the corresponding terms, $Z_1$ and $Z_1 Z_2$, in the BK Hamiltonian are assigned to group $G_{1,BK}$.

\section{Parallel Simulation of Hamiltonians}\label{psh}
Before introducing the parallel simulation circuit, we introduce the unitary transformation $U_n$ that transforms a group of mutually fully-commuting operators $G_n$ into a group of qubit-wise commuting operators $G_n^\prime$. This method was introduced by Bravyi et al.~\cite{bravyi2017tapering}. $U_n$ can be implemented with an evolution circuit depth of $\mathcal{O}(N\log(N))$.

Initially, we identify a set $T = \{T_1, T_2, \cdots, T_N\}$, where the elements are multi-qubit Pauli operators. The elements of $T$ commute with each other and with the elements in $G_n$. Additionally, we find a set $Q = \{\sigma_1, \sigma_2, \cdots, \sigma_N\}$, where $\sigma_j$ is a single-qubit Pauli operator, and the operators in $Q$ commute with each other. The operators in $T$ and $Q$ satisfy the following commutation relations
\begin{equation}
    \begin{split}
       & \{T_i, \sigma_i\} = 0,\\
        & [T_i, \sigma_j]= 0, \; i \neq j.
    \end{split}
\end{equation}
Then, $U_n$ can be expressed in terms of $T_i$ and $\sigma_i$ as 
\begin{equation}
    U_n = \prod_{i=1}^N\frac{1}{\sqrt{2}}(T_i + \sigma_i)=\prod_{i=1}^N V_i.
\end{equation}
For convenience in execution on quantum computers, $V_i$ can also be reexpressed as
\begin{equation}
    V_i = (-i) e^{i\frac{\pi}{4}\sigma_i}e^{i\frac{\pi}{4}T_i}e^{i\frac{\pi}{4}\sigma_i}.
\end{equation}

In an $N$-qubit system, the Pauli weight of $T_i$ is at most $N$. Therefore, the maximum circuit depth required to implement $e^{i\frac{\pi}{4} T_i}$ is $\mathcal{O}(\log(N))$. Since $\sigma_i$ is a single-qubit Pauli operator, it can be implemented using single-qubit rotation gates, so the circuit depth required to implement $V_i$ is $\mathcal{O}(\log(N))$. Given that $U_n$ contains $N$ such $V_i$ operators, the overall circuit depth for $U_n$ is at most $\mathcal{O}(N \log(N))$.

\begin{figure*}[htpb]
    \centering
	\includegraphics[width=0.9\textwidth]{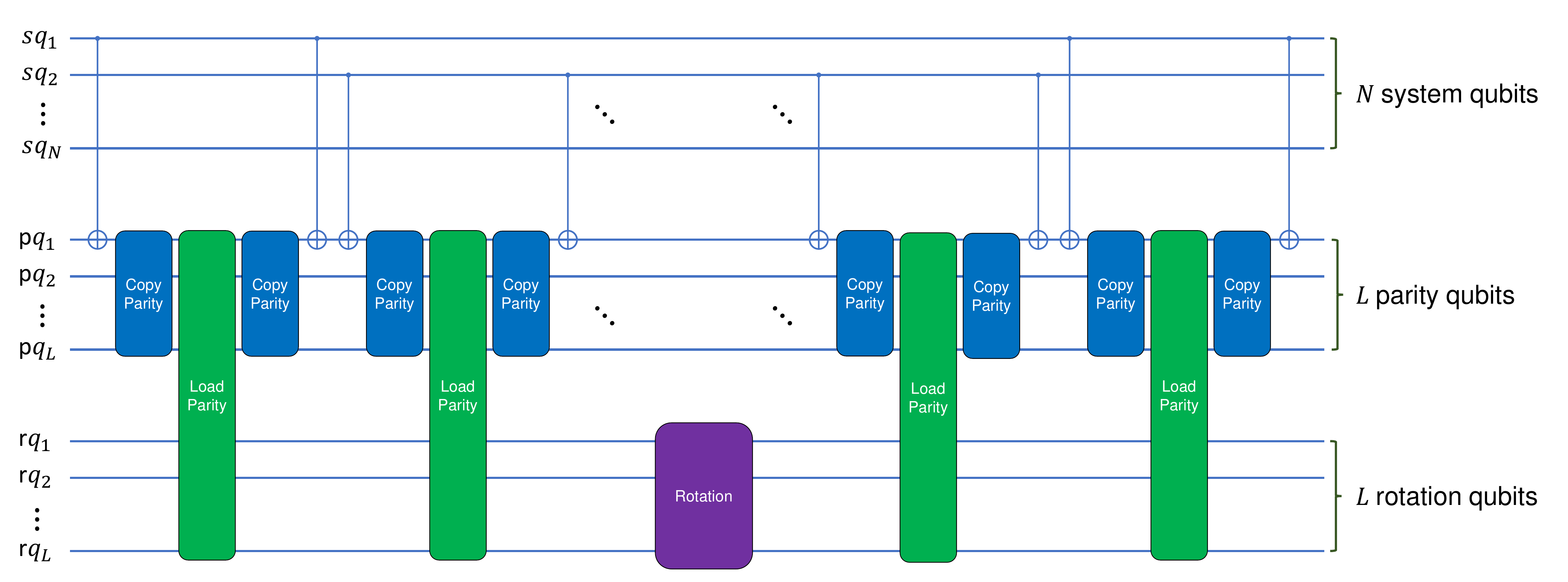} 
    \caption{The circuit for parallel simulation of a qubit-wise commuting group. The copy-parity circuit is used to copy the parity of one qubit to the other parity qubits. The load-parity circuit loads the parity of parity qubits to the corresponding rotation qubits. After loading the parity of all system qubits, rotation gates are applied to the rotation qubits in parallel. $L$ is the number of terms in the  qubit-wise commuting group.}
    \label{fig_pqc}
\end{figure*}

Next, we describe the method for simultaneously evolving a group of qubit-wise commuting operators $G_n^\prime$. First, we use the basis change operations introduced in Sec.~\ref{hsc} to transform these operators into ones composed entirely of Pauli $Z$ operators. We then introduce two sets of auxiliary qubits: one set for parity qubits and another set for rotation qubits. The number of auxiliary qubits in each set is equal to the number of operators in $G_n^\prime$, and each operator corresponds to one rotation qubit. Since $G_n^\prime$ contains at most $\mathcal{O}(N^2)$ operators, the number of auxiliary qubits is also at most $\mathcal{O}(N^2)$.

Then we load the parity of each system qubit onto its corresponding rotation qubit. Specifically, we first copy the parity of the first system qubit to the first parity qubit by applying a CNOT gate. Then, the first parity qubit propagates its parity to the other parity qubits through additional CNOT gates. This step has a circuit depth of $\mathcal{O}(\log(N))$. 

Following this, we selectively load the parity from the parity qubit onto  its corresponding rotation qubit. For instance, if the $j$-th operator $g_j$ in $G_n^\prime$ contains $Z_1$, we apply a CNOT gate between the $j$-th parity qubit and the $j$-th rotation qubit; otherwise, we perform no operation. This part of the circuit has a depth of $\mathcal{O}(1)$. 

\begin{algorithm}[h]\label{alg_huffman_tree}
		\caption{Generating of parallel simulation circuits}
		\KwIn{$qc$ \text{ and } $G = \{H_1,H_2,\cdots, H_L\}$}\tcp{$G$ is a group including qubit-wise commuting Hamiltonian terms.}
		\KwOut{$qc$} \tcp{$qc$ is the quantum circuit.}
  \textbf{Initialize}: $\ket{sq} = \ket{\Psi}, \ket{pq} = \ket{0}^{\bigotimes L}, \ket{rq} = \ket{0}^{\bigotimes L}$ \;\tcp{$sq$, $pq$, and $rq$ are sets of system, parity, and rotation qubits, respectively}
        \textbf{Define Function: Copy\_parity($qc,j,N_p, sq, pq$)}\{
        
        $qc$.CNOT($sq[j]$, $pq[1]$)\;\tcp{$sq[j]$ is the $j$th system qubit}
        
        $I = 1$\;
        \While{$I < N_p$}{
          \For{$i=1$ \text{to} $I$}
        {
        $qc$.CNOT($pq[i], pq[i+I]$);\tcp{Insert a CNOT gate.}
        
        \If{$i+I == L$}
        {break\;}
        }
        $I = I * 2$\;
        }
        
        \Return $qc$\;
        \}

        \textbf{Define Function: Load\_parity($qc,G, sq, pq, rq$)}\{
        
        \For{$j=1$ to $N$}{

        $N_p = 0$\;
         \For{$i=1$ \text{to} $L$}
        {
        \If{$X_j$ in $G[i]$ or $Y_j$ in $G[i]$ or $Z_j$ in $G[i]$}
        {  
           $N_p$ += 1\;
        }
        }
        
        $qc$ = Copy\_parity($qc, j, N_p, sq, pq$)\;\tcp{copy the parity of the $j$-th system to $N_p$ parity qubits }
         \For{$i=1$ \text{to} $L$}
        {
        $N_p = 0$\;
        \If{$X_j$ in $G[i]$ or $Y_j$ in $G[i]$ or $Z_j$ in $G[i]$}
        {  
            $N_p$ += 1\;
            $qc$.CNOT($pq[N_p], rq[i]$)\;
        }
        
        }
         $qc$ = Copy\_parity($qc, j, N_p, sq, pq$)\;\tcp{recover the $N_p$ parity qubits }
        
        }
         \Return qc\;
        \}

        $qc$ = Load\_parity($qc,G, sq, pq, rq$)\;
         \For{$i=1$\text{to} $L$}
         {
         $qc$.RZ($rq[i]$, $H_i$.cof)\tcp{$H_i$.cof is the coefficient of the Hamiltonian term $H_i $}
         }
         $qc$ = Load\_parity($qc,G, sq, pq, rq$)\;

\end{algorithm}

Subsequently, the parity of the first qubit is removed from the parity qubits by repeating the parity copying operation for the first qubit. This procedure is repeated until the parity of all system qubits is selectively copied onto the rotation qubits. Loading the parity of one system qubit onto a rotation qubit requires a circuit depth of $\mathcal{O}(\log(N))$. Thus, loading the parity of $N$ system qubits requires a total circuit depth of $\mathcal{O}(N \log(N))$. 

After loading the parity, $R_Z(\theta_j)$ rotations are applied simultaneously to the rotation qubits, with the rotation angle depending on the coefficient of the corresponding Hamiltonian term in the group. After the rotations, the parity of each system qubit is subtracted from the rotation qubits using the same procedure as the parity loading process. Finally, we perform the basis change operations followed by the application of $U_n^\dagger$. The Algorithm~(\ref{alg_huffman_tree}) shows how to generate the parallel simulation circuit, and the parallel simulation circuit is shown in Figure \ref{fig_pqc}. The copy-parity circuit copies the parity of a system into $L$ copies, enabling the simultaneous loading of the parity of a system qubit onto $L$ rotation qubits. The load-parity circuit then loads these $L$ copies of the parity onto the rotation qubits simultaneously.

Thus, the evolution of $G_n$ requires a circuit depth of $\mathcal{O}(N \log(N))$. For a Hamiltonian with $\mathcal{O}(N^2)$ groups, the total circuit depth for evolving the Hamiltonian is $\mathcal{O}(N^3 \log(N))$. In contrast, without grouping and parallel evolution, the circuit depth for a single Trotter step of evolution would be $\mathcal{O}(N^4 \log(N))$. Compared to the original scheme, our method reduces the complexity by a factor of $N$.
    
\section{Simultaneous Measurement by Grouping}\label{smg}
First, we consider the expectation value measurements of the Pauli operators in the $G_1$ group. Since all Pauli strings in this group are composed of Pauli $Z$ or $I$ operators, we can directly perform Z-basis measurements on all the qubits.

\begin{figure}[h]
    \centering
	\includegraphics[width=0.35\textwidth]{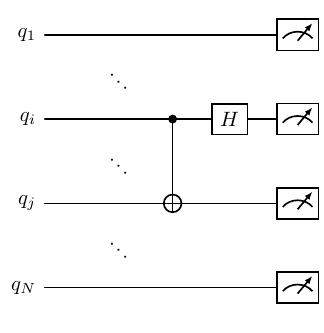} 
    \caption{The circuit to measure expectation values of Pauli strings in $G_2(a,b)$.}
    \label{fig_g2_circ}
\end{figure}

To measure $G_2(i, j)$, we use the Bell measurement circuit. The Bell measurement circuit is applied to the qubits on which Pauli $X$ or $Y$ operators act. Note that
\begin{equation}
    \begin{split}
        &H_i \text{CNOT}(i, j) X_i X_j \text{CNOT}(i, j) H_i = Z_i, \\
         &H_i \text{CNOT}(i, j) Y_i Y_j \text{CNOT}(i, j) H_i = -Z_i Z_j,
    \end{split}
\end{equation}
where $H_i$ denotes the Hadamard gate applied to the $i$-th qubit, and $\text{CNOT}(i, j)$ represents a controlled-not gate with the $i$-th qubit as the control qubit and the $j$-th qubit as the target qubit. Therefore, we can add Bell measurement circuits to the qubits where $XX$ and $YY$ operators appear, while performing Z-basis measurements directly on the remaining qubits, as shown in Figure~\ref{fig_g2_circ}.

For the measurements of operators in groups $G_3$ to $G_8$, no simple auxiliary circuits are currently available. Therefore, we will use the conventional approach for measuring a group of fully-commuting operators. Specifically, we will apply the $U_n$ transformation introduced in the previous section to convert a group of fully-commuting operators into a group of qubit-wise commuting operators. Subsequently, basis change operations will transform these operators into ones composed entirely of Pauli $Z$ operators. Finally, we measure all system qubits to complete the measurements of a group  $G_3$ to $G_8$.

If we consider the number of circuits required for measuring a single operator or a group of operators as a constant, then the grouping measurement approach only requires $\mathcal{O}(N^2)$ circuits. In contrast, measuring each operator individually would require $\mathcal{O}(N^4)$ circuits to estimate the expectation value of the molecular Hamiltonian. This represents a quadratic reduction in complexity compared to the original method.

Although the auxiliary circuit $U_n$ introduces additional gates into the total circuits, its depth is relatively small compared to the entire circuit depth. For example, the Hamiltonian evolution circuit still requires a depth of at least $\order{N^3 \log(N) M}$ with our method. In contrast, the depth of the $U_n$ circuit is only $\order{N \log(N)}$, which is much less than the depth of the Hamiltonian evolution circuit. Even in the VQE algorithm, commonly used ansatze, such as the hardware-efficient ansatze, have a circuit depth of $\mathcal{O}(N^2)$, which is much greater than the depth of the auxiliary circuit $U_n$. Furthermore, $U_n$ can be regarded as the evolution circuit for a Hamiltonian with $N$ terms. Considering that the original Hamiltonian has $\order{N^4}$ terms, the addition of a much shallower auxiliary circuit with a similar structure should not significantly increase the impact of noise. Interestingly, if we combine the parallel simulation and the simultaneous measurement schemes, when measuring the expectation value of $G_n$, placing $G_n$ at the end of the evolution circuit can eliminate the circuits of $U_n^\dagger$ and $U_n$.

In the NISQ era, noise levels are relatively high, limiting us to computing small-scale systems. In such cases, the circuits are typically shallow, and the number of Hamiltonian terms is not large, indicating that time is not the primary bottleneck. The need for auxiliary circuits in grouping measurement and parallel Hamiltonian simulation leads to an increase in circuit depth. Nevertheless, our scheme can easily degrade to the (near-) QWC grouping scheme, which either does not require auxiliary circuits or only requires auxiliary circuits with very shallow depths of $\order{1}$. Under the degenerate grouping scheme, the number of groups is $\order{N^3}$, which provides an advantage over other works. This is because other QWC grouping schemes still require $\order{N^4}$ groups, with no reduction in complexity, resulting only in a reduction of the constant factor.
For example, for the four-body term $\widehat{Y_i Y_j Y_k Y_l}$, we can group it as follows
\begin{equation}
   G(a,b,m_2) = \{\widehat{Y_i Y_j Y_k Y_l} | i = a, j = b, \frac{k+l}{2} = m_2\}.
\end{equation}
In this grouping scheme, the auxiliary circuits for simultaneous measurement and parallel evolution of a group of operators consist of a layer of CNOT gates.

However, in the fault-tolerant quantum era, we will be able to simulate much larger systems. At that stage, due to the high time complexity of many quantum algorithms for quantum chemistry, time becomes the main bottleneck. Our parallel simulation and simultaneous measurement schemes can significantly reduce computational time. Although our approach is effective in the NISQ era, it will offer even greater computational benefits in the fault-tolerant quantum computing era.

\section{Numerical Test Results}\label{ntr}
We calculated the number of Hamiltonian terms in the second quantization representation for molecular systems with different numbers of orbitals, as well as the number of groups formed using our grouping strategy. The results are shown in Table~\ref{tableofgroups}.

\begin{table}[!htbp]
		\centering
		\caption{ The total number of Hamiltonian terms and the number of groups. All Hamiltonians are calculated in the STO-3G basis. The coefficient threshold for Hamiltonian terms is set to $10^{-8}$.\label{tableofgroups}}
	\begin{tabular}[c]{llll}
			\hline \hline
			Molecules			&Qubits			&Terms 	 &Groups     	\\
			\hline
			
			LiH					&12		& 631		 & 191          	\\
			H$_2$O				&14		& 1930	 & 457        \\
			NH$_3$				&16		& 5281	 & 823         \\
			Mg					&18 	&2092		 &615         	\\
			N$_2$				&20		&2951    	 &809         	\\
                C$_2$H$_2$		   	&24    	&18957	 &2263        	\\
			CO$_2$				&30	   & 30534	 &3319      	\\
                C$_2$H$_6$			&32    	&90993	 &4759         	\\
			Cl$_2$				&36	   &34583	 &4385     	\\
			HNO$_3$				&42	  & 161224   &8395       \\
			CH$_3$COOH	        &48     &422397	   & 11873     \\	
			\hline \hline
		\end{tabular}
	\end{table}

It can be observed that the actual number of Hamiltonian terms is less than $N^4$. In practice, the number of terms is approximately one-tenth of $N^4$. This reduction is due to molecular symmetries such as parity conservation and possible spatial symmetries, resulting in some Hamiltonian terms having zero coefficients.

\begin{figure}[h]
		\centering

       \subfigure[]{
    			\includegraphics[width=0.38\textwidth]{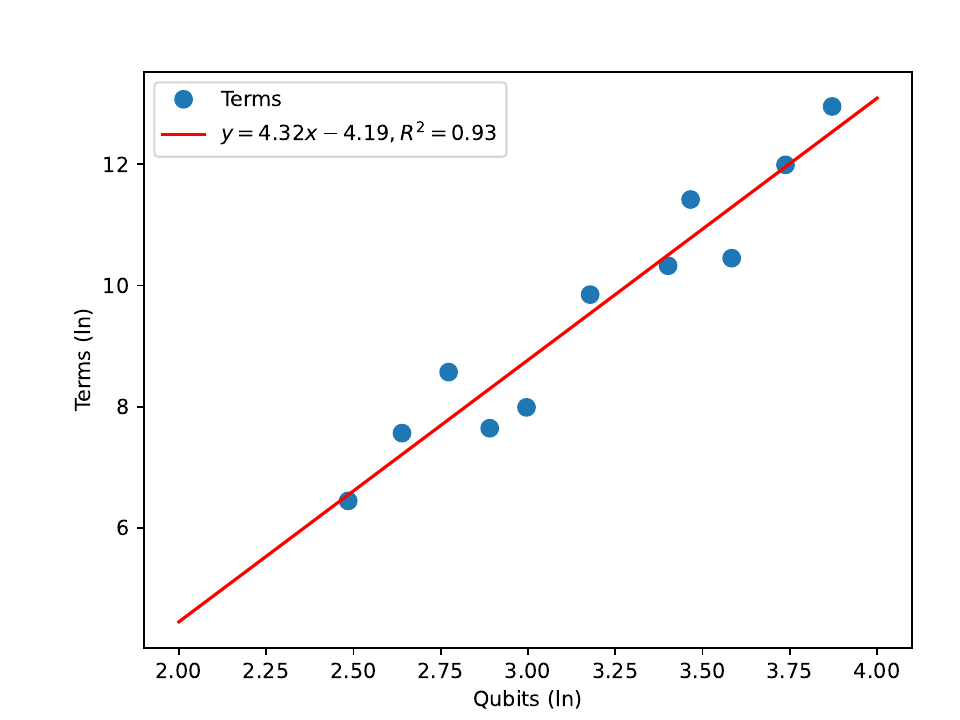}\label{log_qubit_term}    }
                  \subfigure[]{
                	\includegraphics[width=0.38\textwidth]{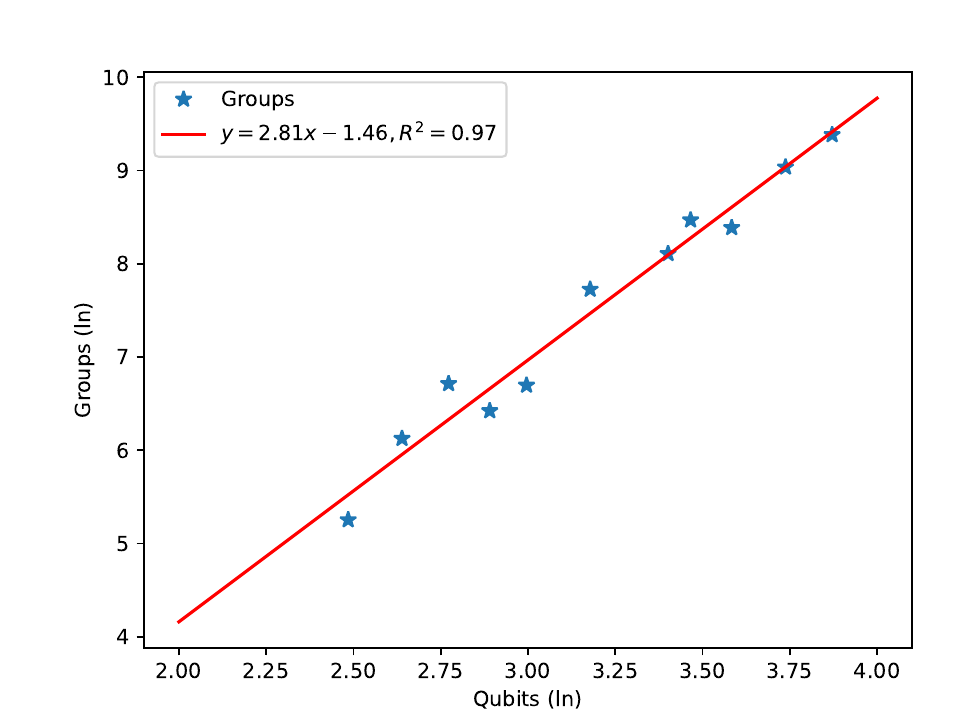}\label{log_qubit_group}    }
		\caption{ Dependencies of the total number of terms and the number of groups on the number of qubits for the molecular Hamiltonian in Table~\ref{tableofgroups}. a) The total number of Hamiltonian terms along with the corresponding fitted curve. b) The number of commuting groups and their fitted curve. }
        \label{fig_groups}
	\end{figure}
\begin{figure}
    \centering
    \includegraphics[width=0.85\linewidth]{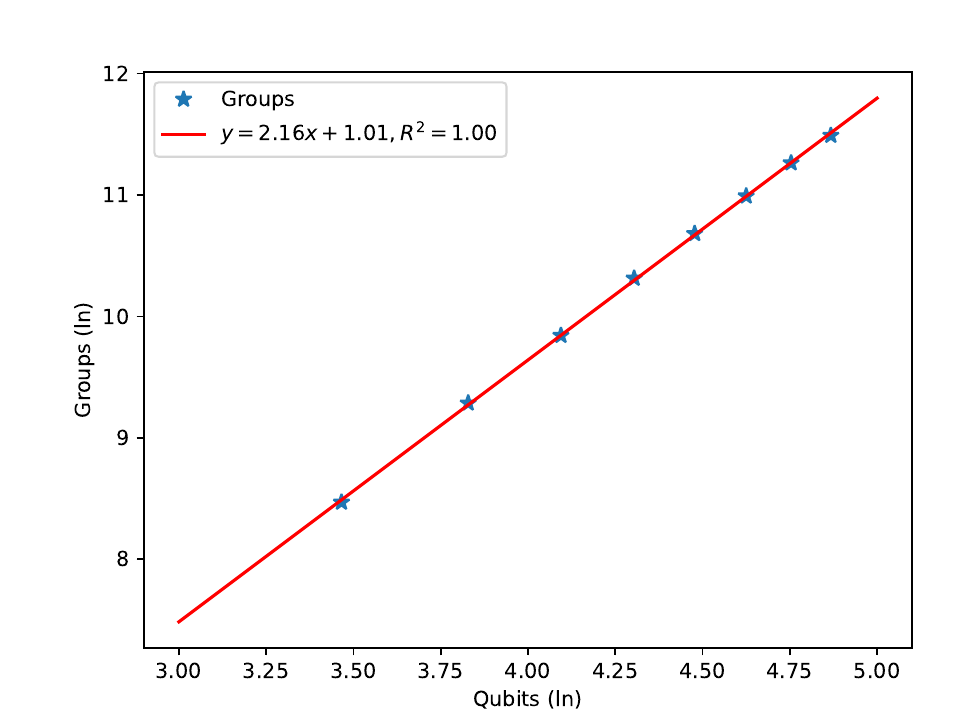}
    \caption{Dependencies of the total number of terms and the number of groups on the number of qubits for the molecular Hamiltonians of a series of alkanes ranging from C$_2$H$_6$ to C$_9$H$_{20}$. }
    \label{fig_alk_groups}
\end{figure}
We also analyzed the relationship between the number of terms and the number of groups with respect to the number of qubits using a logarithmic scale. The results are shown in Figure~\ref{fig_groups}.

\begin{figure}
    \centering
    \includegraphics[width=0.85\linewidth]{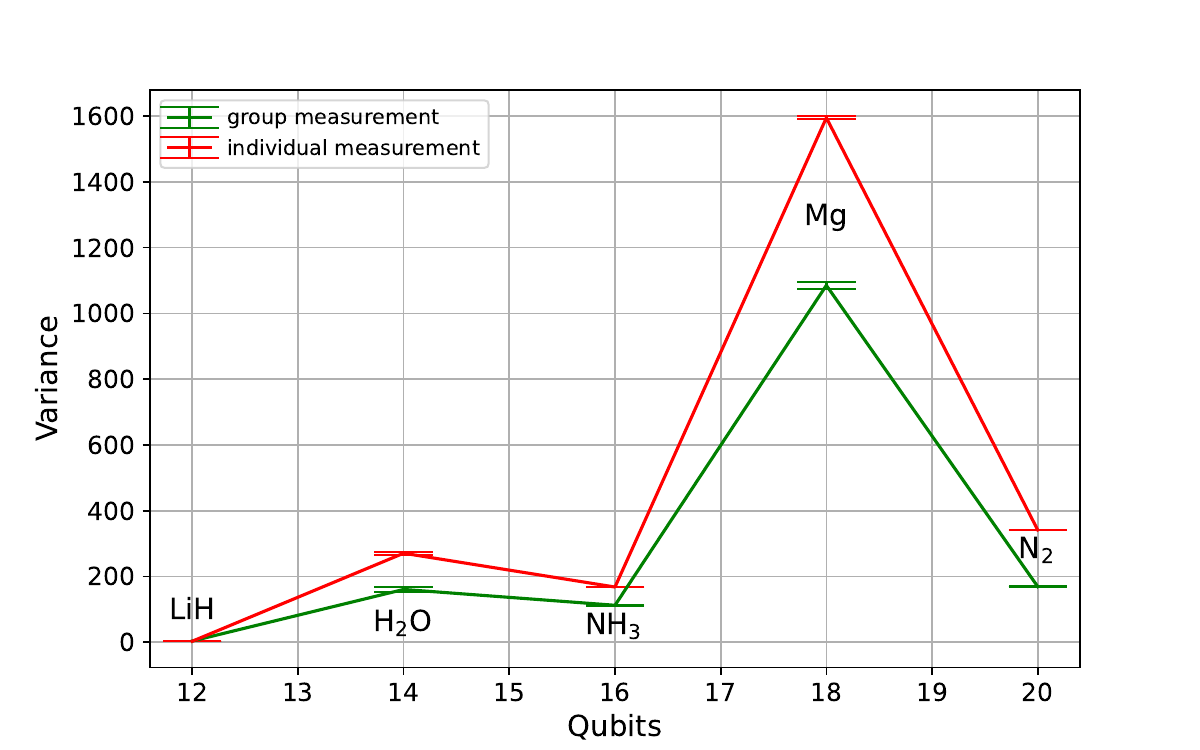}
    \caption{Variance of individual measurement scheme (red line) and grouping measurement strategy (green line). }
    \label{fig_var}
\end{figure}

According to the fitted curve, the number of groups scales as $\mathcal{O}(N^{2.81})$. Although the fitted scaling is greater than $\mathcal{O}(N^{2})$, our grouping strategy theoretically guarantees that the complexity of the number of groups is $\mathcal{O}(N^{2})$. The fitted scaling greater than $\mathcal{O}(N^{2})$ is due to the fact that when the molecule is relatively small, many Hamiltonian terms are eliminated due to symmetry, which reduces the efficiency of our method.

Considering the reasons mentioned above, we additionally tested a series of alkanes (from C$_2$H$_6$ to C$_9$H$_{20}$), and the result is shown in Figure~\ref{fig_alk_groups}. These molecules, with similar structures, are more suitable for fitting.  Due to less symmetry and more orbitals, the estimated scaling of the number of groups is reduced to $\mathcal{O}(N^{2.16})$, which closely approximates the theoretical scaling of  $\mathcal{O}(N^{2})$.

We also calculated the total variance for both individual terms and groups. Total variance determines the number of shots needed to measure the expectation value of a term or a group. The variance was calculated using random states that satisfy particle number conservation, averaged over the samples. The results are depicted in Figure \ref{fig_var}. This shows that the total variance of all groups is smaller than the total variance of all terms. Thus, measuring the expectation value of a group requires fewer shots than measuring an individual term.

\section{CONCLUSION AND DISCUSSION}\label{cad}

In this work, we developed a strategy to group the terms in a molecular Hamiltonian, effectively reducing the number of original $\mathcal{O}(N^4)$ terms to $\mathcal{O}(N^2)$ groups, where the terms within each group commute with each other. This grouping enabled a refinement of the Hamiltonian evolution scheme based on Trotter decomposition, enabling parallel evolution of commutative terms within groups, thereby decreasing the circuit depth complexity by a factor of $N$. Our grouping strategy also enhances the efficiency of Hamiltonian expectation value measurements, achieving an order of $N$ improvement over the current best grouping measurement techniques. Moreover, we observed that the shots required per group are fewer than those for measuring each term individually. This optimization reduces both the number of groups and the number of measurements per group. Collectively, our approach not only diminishes the execution time for each circuit operation but also significantly decreases the total number of circuit executions, resulting in an overall computational time improvement of at least $\mathcal{O}(N^3)$.

Both the parallel evolution of Hamiltonian and the grouping measurement require auxiliary circuits, denoted as $U_n$, which contribute additional circuit depth. However, compared to the original algorithms (e.g., VQE and QSP), the increase in circuit depth due to $U_n$ is relatively small and can be considered negligible. The $U_n$ circuit is designed to transform a set of arbitrarily full-commuting operators into qubit-wise commuting operators. Given that the operators in $G_n$ after our grouping method have a certain structure, there may exist simpler transformations to achieve this purpose.

	\acknowledgments
	 The numerical calculations in this paper were performed on the supercomputer system at the China Supercomputer Center of the University of Science and Technology of China.
	This work was supported by the National Key Research and Development Program of China (Grant No. 2024YFB4504101), the National Natural Science Foundation of China (Grant No. 22303022), and the Anhui Province Innovation Plan for Science and Technology (Grant No. 202423r06050002).

    \appendix

\section{More discussion on the grouping scheme}\label{dis_scheme}
We have already introduced our grouping scheme in Sec.~\ref{gph}. Now, we prove that the Hamiltonian terms in \( G_n \) commute with each other. In the proof, we will use the following basic commutation relations 
\begin{equation}
    \begin{split}
        [\sigma_i, \sigma_i] &= 0,\\
        \{\sigma_i, \sigma^\prime_i\} &= 0, \quad \sigma \neq \sigma^\prime,\\
        [\sigma_i, \sigma^\prime_j] &= 0, \quad i \neq j, 
    \end{split}
\end{equation}
where $\sigma \in \{X, Y, Z\}.$

The terms in \( G_1 \) are obviously mutually commuting.  It is also straightforward to show that the terms in \( G_2(a,b) \) commute with each other by using the relation \( [X_aX_b, Y_aY_b] = 0 \). 
The more challenging part is proving the commutativity of groups that contain four-body terms. Instead of proving this for all such groups, we will focus on proving that the terms in \( G_8(m_1, m_2) \) commute with each other. For other groups, the proof follows a similar approach.  

Consider any two elements, P1 ($Y_iZ_{i+1}\cdots Z_{j-1}X_j X_kZ_{k+1}\cdots Z_{l-1}Y_l$) and P2 ($Y_{i^\prime} Z_{{i^\prime} +1}\cdots Z_{j^\prime -1}X_{j^\prime}  X_{k^\prime}Z_{k^\prime +1}\cdots Z_{l^\prime-1}Y_{l^\prime}$), from \( G_8 \). We aim to show that they commute. For convenience, we split P1 into LP1 ($Y_iZ_{i+1}\cdots Z_{j-1}X_j$) and RP1 ($X_kZ_{k+1}\cdots Z_{l-1}Y_l$), and P2 into LP2 ($Y_{i^\prime} Z_{{i^\prime} +1}\cdots Z_{j^\prime -1}X_{j^\prime}$) and RP2 ($X_{k^\prime}Z_{k^\prime +1}\cdots Z_{l^\prime-1}Y_{l^\prime}$).  

\begin{figure}[h]
		\centering
	
       \subfigure[]{
    			\includegraphics[width=0.45\textwidth]{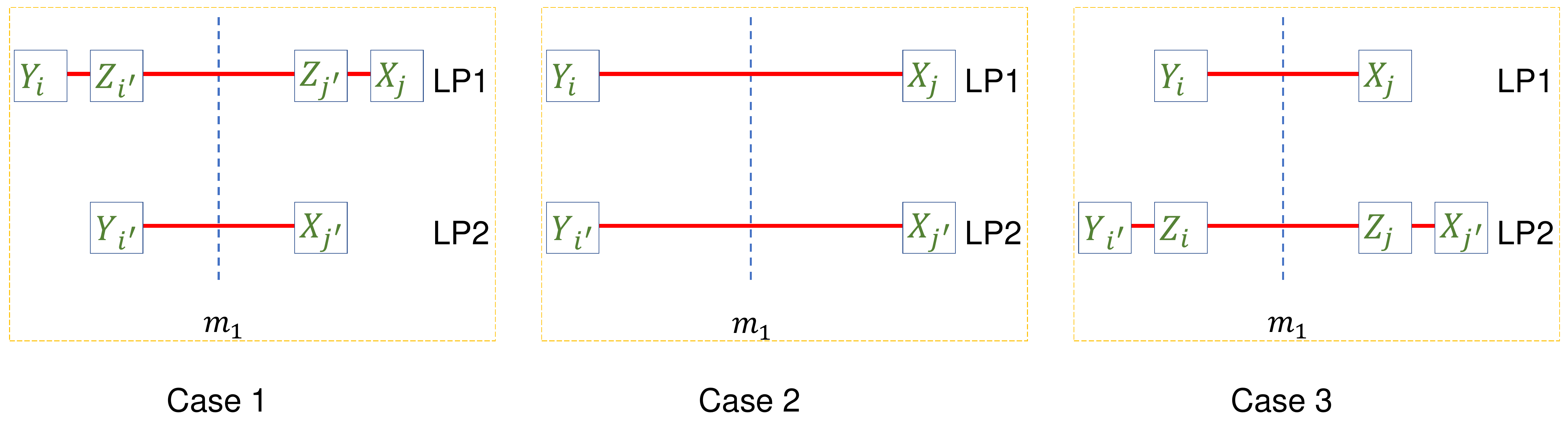}\label{fig_LP_LP}    }\\
                  \subfigure[]{
                	\includegraphics[width=0.45\textwidth]{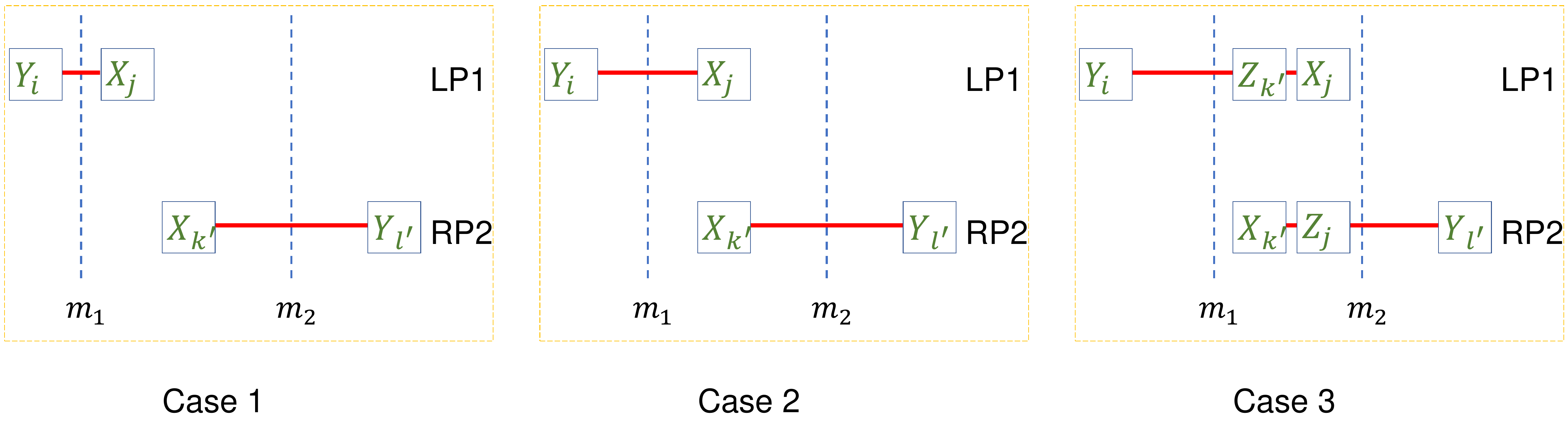}\label{fig_LP_RP}    }
		\caption{ The relationship of Pauli substrings in P1 and P2. a) The relationship of LP1 and LP2. b) The relationship of LP1 and RP2. }
        \label{fig_commu}
	\end{figure}

First, we prove that LP1 and LP2 commute. Their relationship is illustrated in Figure~\ref{fig_LP_LP}. In Case 1, since \( \{Z_i^\prime, Y_i^\prime\} = 0 \) and \( \{Z_j^\prime, X_j^\prime\} = 0 \), it follows that LP1 and LP2 commute. In the remaining two cases, tt is easy to see that LP1 and LP2 also commute.  Next, we prove that LP1 and RP2 commute. The relationship of LP1 and RP2 is shown in Figure~\ref{fig_LP_RP}. In Cases 1 and 2, it is straightforward to see that LP1 and RP2 commute. In Case 3, using the relations \( \{Z_k^\prime, X_k^\prime\} = 0 \) and \( \{X_j, Z_j\} = 0 \), we can demonstrate that LP1 and RP2 commute.  

In fact, in all cases, there are always an even number of pairs of single-qubit Pauli operators that anticommute within LP1 and LP2 (or RP2). Therefore, LP1 and LP2 (or RP2) always commute. Consequently, LP1 commutes with P2. Since LP1 and RP1 share a similar structure, it naturally follows that RP1 also commutes P2. Since LP1 and RP1 both commute with P2, we conclude that P1 and P2 commute. 

To facilitate the implementation of our grouping scheme for readers, we provide a simple and efficient algorithm for grouping, Algorithm~\ref{alg_grouping}. The core idea is to determine the type of each term and compute two indices based on this type. For example, consider the term $X_1X_2 X_4Z5X6$. Its Pauli type is \( XXXX \), and the two indices are \( \frac{j+k}{2} = 3 \) and \( (l-k) - (j-l) = 1 \). The two indices, along with the Pauli type, form a ``label." We use a dictionary (a fundamental data structure in Python and C++) to store the grouping results. The dictionary's keys correspond to group labels, and the values are sets that store the Hamiltonian terms belonging to each group. In special cases, some groups do not require the two indices. For instance, for \( G_1 \), we fix the two indices to 0 for uniformity. Since this group contains different types of terms, we use ``I" to represent the type of this group.

\begin{algorithm}[h]\label{alg_grouping}
		\caption{The algorithm for Hamiltonian grouping}
		\KwIn{Hamitonian $H = \{h_1H_1, \cdots, h_LH_L\}$}
		\KwOut{$G_{\text{dict}}$} \tcp{$G_{\text{dict}}$ is a dictionary that stores the grouping results.}
         \textbf{Initialize}: $G_{\text{dict}} = \{\}$\;\tcp{The keys of $G_{\text{dict}}$ represent the labels of groups, and the values of $G_{\text{dict}}$ are empty sets to store the Hamiltonian terms.}
        \textbf{Define Function: Get\_Pauli\_type($H_m$)}\{
        
        $\cdots$;
        \tcp{Obtain the Pauli type of $H_m$ according to the summary in Table~\ref{tableofPauliClass}.}
        \Return $type$\;
        \}
        
         \For{$m=1$ \text{to} $L$}
        {
        $type$ = Get\_Pauli\_type($H_m$)\;
        \If{type in [``I", ``Z", ``ZZ"]}
        {  
            $index1$ = $index2$ = 0\;
            $label$ = ``I" + Str($index1$) + Str($index2$)\;\tcp{Str is a function that converts a number to a string.}
        }
        \If{type in [``ZXX", ``ZYY", ``XZX'', ``YZY", ``XXZ", ``YYZ"]}
        {  
            $index1$ = $i$, $index2$ = $j$\;\tcp{The $i$ and $j$ are the indices of the first and second Pauli $X$ ($Y$) operators in $H_m$, respectively.}
            $label$ = ``AA" + Str($index1$) + Str($index2$)\;
        }
        \If{type in [``XXXX", ``YYYY", ``XXYY", ``YYXX"]}
        {  
            $index1$ = $\frac{j+k}{2}$, $index2$ = $(l-k)-(j-l)$\;\tcp{The $j$ and $k$ are the indices of the third and fourth Pauli $X$ ($Y$) operators in $H_m$, respectively.}
            $label$ = $type$ + Str($index1$) + Str($index2$)\;
        }
    \If{type in ["XYYX", "YXXY"]}
        {  
            $index1$ = $\frac{i+j}{2}$, $index2$ = $\frac{k+l}{2}$\;
            $label$ = $type$ + Str($index1$) + Str($index2$)\;
        }
        $G_{\text{dict}}$[$label$].insert[$H_m$]\;\tcp{Insert $H_m$ into the set with the key = $label$.}

        }
        
         \Return $G_{\text{dict}}$ \;
	
\end{algorithm}

\section{More discussion on the parallel simulation of Hamiltonians}\label{dis_psh}
In Sec.\ref{hsc}, we introduced Hamiltonian evolution circuits based on Trotter decomposition. Building upon these circuits, we then proposed the Hamiltonian parallel evolution algorithm. Here, we will provide a step-by-step explanation of how this parallel evolution algorithm works.

First, consider the evolution operator of a multi-qubit Pauli operator \( P \), \( e^{-iPt} \), where \( t \) is the evolution time. Since single-qubit Pauli \( X \) and \( Y \) operators in \( P \) can be transformed into Pauli \( Z \) operators using a simple basis-change gate, we focus here on the case where \( P \) consists solely of Pauli \( Z \) operators.  

If \( P \) is a single-qubit Pauli \( Z \) operator, its evolution operator \( e^{-iPt} \) can be implemented using a single-qubit \( R_Z(\theta, k) \) gate, where \( \theta=\frac{t}{2} \) is the rotation angle, and \( k \) denotes the qubit on which the \( R_Z \) gate is applied.  

For a two-qubit operator, it cannot be directly implemented using a single quantum gate. Utilizing the linearity of quantum mechanics, we can just analysis the effect of the operator on a computational basis state. Suppose \( P = Z_j Z_k \), then applying the evolution operator to a computational basis state \( \ket{q_1, q_2, \dots, q_N} \) results in  
\begin{equation}
    e^{-iZ_j Z_k t} \ket{q_1, q_2, \dots, q_N} \rightarrow e^{-it(-1)^{q_j + q_k}} \ket{q_1, q_2, \dots, q_N}.
\end{equation}
As we can see, this operator does not change the state of individual qubits; instead, it introduces a phase factor, whose sign depends on the parity of \( q_j + q_k \). Notably,the \( R_Z \) gate also does not alter the qubit state but introduces a phase, whose sign depends on the parity of the qubit on which \( R_Z \) is applied. Exploiting this property, we can implement the evolution operator of \( P \) using \( CNOT \) and \( R_Z \) gates.  

The basic idea is as follows. First, Apply a \( CNOT(j, k) \) gate to load the parity of the \( j \)-th qubit to the \( k \)-th qubit. Then Apply an \( R_Z \) gate on the \( k \)-th qubit to introduce the phase \( e^{-i(-1)^{q_j + q_k} t} \). At last, Apply another \( CNOT(j, k) \) gate to restore the state of the \( j \)-th qubit.This process can be expressed as
\begin{equation}
    \begin{split}
     &CNOT(j,k)R_Z(\theta, k)CNOT(j,k)\ket{\cdots, q_j, \cdots, q_k, \cdots} \\
     \rightarrow &CNOT(j,k)R_Z(\theta, k)\ket{\cdots, q_j, \cdots, q_j+q_k, \cdots}\\
     \rightarrow &CNOT(j,k)e^{-it(-1)^{q_j + q_k}}\ket{\cdots, q_j, \cdots, q_j+q_k, \cdots}\\
     \rightarrow &e^{-it(-1)^{q_j + q_k}}\ket{\cdots, q_j, \cdots, q_k, \cdots}.
\end{split}
\end{equation}
Note that \( q_j + q_k \) actually represents \( (q_j + q_k) \mod 2 \).  

Additionally, we can implement this evolution operator using an auxiliary qubit initialized to \( \ket{0} \). The basic idea is to use two \( CNOT \) gates to load the parity of the \( j \)-th and \( k \)-th qubits onto the auxiliary qubit. Then, applying an \( R_Z \) gate on the auxiliary qubit generates the corresponding phase. Finally, two more \( CNOT \) gates restore the auxiliary qubit to \( \ket{0} \).  

This method generalizes naturally to cases where \( P \) acts on an arbitrary number of qubits. To formalize this, we define a set \( Q(P) \), which records the qubits on which $P$ acts
\begin{equation}
    Q(P) = \{ j \in [1,N] \mid Z_j \text{ appears in } P \}.
\end{equation}
The circuit for implementing \( e^{-iPt} \) can be conveniently expressed as  
\begin{widetext}
\begin{equation}
    \begin{split}
    &\left(\prod_{j \in Q(P)}CNOT(j,rq)\right) R_Z(\theta, rq) \left(\prod_{j \in Q(P)}CNOT(j,rq)\right) \ket{q_1, \cdots, q_N}\ket{0} \\
    \rightarrow &\left(\prod_{j \in Q(P)}CNOT(j,rq)\right) R_Z(\theta, rq) \ket{q_1, \cdots, q_N}\ket{\sum_{j \in Q(P)}q_j}\\
    \rightarrow &\left(\prod_{j \in Q(P)}CNOT(j,rq)\right) e^{-it(-1)^{\sum_{j \in Q(P)}q_j}}\ket{q_1, \cdots, q_N}\ket{\sum_{j \in Q(P)}q_j}\\
    \rightarrow & e^{-it(-1)^{\sum_{j \in Q(P)}q_j}}\ket{q_1, \cdots, q_N}\ket{0}.
\end{split}
\end{equation}
\end{widetext}

Now, we consider the evolution of a set of mutually commuting operators \( G = \{ P_l, \dots, P_L \} \). For simplicity, we assume that each \( P_l \in G \) consists solely of Pauli \( Z \) operators. Since these operators commute, the order of evolution does not affect the result. Thus, we omit specifying the evolution order in the following equations.  

The action of the evolution operators for a set of $G$ on a computational basis state is  
\begin{equation}
    \left(\prod_l e^{-iP_l t_l} \right) \ket{\mathbf{q}} = \left(\prod_l e^{-it_l(\sum_{j \in Q(P_l)}q_j) } \right) \ket{\mathbf{q}},
\end{equation}
where $\ket{\mathbf{q}}$ is $\ket{q_1, \cdots, q_N}$.
As we can see, the effect of these evolution operators is simply the production of a set of phase factors. Now, we need to prove that the circuit in Figure~\ref{fig_pqc}, when applied to a computational basis state, produces the same production of phase factors.  

To implement these evolution operators in parallel, we introduce two sets of auxiliary qubits: parity qubits and rotation qubits. The goal is to apply \( R_Z \) gates on the rotation qubits. However, before applying these gates, we need to load the parity of the relevant system qubits onto the rotation qubits. This is done using CNOT gates.  

Since multiple CNOT gates acting on the same qubit cannot be executed simultaneously, loading the parity of a single system qubit onto multiple rotation qubits requires multiple layers of CNOT gates. To accelerate this parity-loading process, we introduce parity qubits. The purpose of the parity qubits is to efficiently copy the parity of a system qubit using a shallow circuit. Once the parity of a system qubit is stored in the parity qubits, we load parity of parity qubits  onto the corresponding rotation qubits in a single layer of CNOT gates.  

After sequentially loading the parity of the system qubits onto the rotation qubits, we simultaneously apply a set of rotation operations on the rotation qubits. Once the rotations are completed, the system qubits are restored to \( \ket{0} \) using the same parity-loading circuit.

Therefore, the effect of the circuit in Figure~\ref{fig_pqc} can be expressed as
\begin{widetext}
    \begin{equation}
\begin{split}
    &(\prod_{j=1}^N CNOT(\text{sq}_j, \text{pq}_1)U_{\text{CP}}U_{\text{LP}}(j)U_{\text{CP}}CNOT(\text{sq}_j, \text{pq}_1)) \times\\
    &(\prod_{l=1}^LR_Z(\theta_l, \text{rq}_l)(\prod_{j=1}^N CNOT(\text{sq}_j, \text{pq}_1)U_{\text{CP}}U_{\text{LP}}(j)U_{\text{CP}}CNOT(\text{sq}_j, \text{pq}_1))\ket{\mathbf{q}}_{\text{sq}}\ket{\mathbf{0}}_{\text{pq}}\ket{\mathbf{0}}_{\text{rq}}\\
    \rightarrow &   (\prod_{j=1}^N CNOT(\text{sq}_j, \text{pq}_1)U_{\text{CP}}U_{\text{LP}}(j)U_{\text{CP}}CNOT(\text{sq}_j, \text{pq}_1)) (\prod_{l=1}^LR_Z(\theta_l, \text{rq}_l))\ket{\mathbf{q}}_{\text{sq}}\ket{\mathbf{0}}_{\text{pq}}\ket{\sum_{k \in Q(P_1)}q_k,\cdots,\sum_{k \in Q(P_L)}q_k }_{\text{rq}} \\
    \rightarrow &   (\prod_{j=1}^N CNOT(\text{sq}_j, \text{pq}_1)U_{\text{CP}}U_{\text{LP}}(j)U_{\text{CP}}CNOT(\text{sq}_j, \text{pq}_1)) (\prod_{l=1}^L e^{-it_l\sum_{k \in Q(P_l)}q_k})\ket{\mathbf{q}}_{\text{sq}}\ket{\mathbf{0}}_{\text{pq}}\ket{\sum_{k \in Q(P_1)}q_k,\cdots,\sum_{k \in Q(P_L)}q_k }_{\text{rq}} \\
    \rightarrow &   (\prod_{l=1}^L e^{-it_l\sum_{k \in Q(P_l)}q_k})\ket{\mathbf{q}}_{\text{sq}}\ket{\mathbf{0}}_{\text{pq}}\ket{\mathbf{0}}_{\text{rq}} \\
\end{split}
\end{equation}
\end{widetext}
where \( U_{\text{CP}} \) is the copy-parity circuit shown in Figure~\ref{fig_pqc}. After copying the parity of the \( j \)-th system qubit to the first parity qubit, the \( U_{\text{CP}} \) circuit further propagates this parity to the remaining parity qubits
\begin{equation}
    \begin{split}
    &U_{\textbf{CP}}CNOT(\text{sq}_j, \text{pq}_1)\ket{q_1,\cdots, q_N}_{\text{sq}}\ket{\mathbf{0}}_{\text{pq}}\ket{\mathbf{0}}_{\text{rq}}\\
    \rightarrow& U_{\textbf{CP}}\ket{q_1,\cdots, q_N}_{\text{sq}}\ket{q_j, 0, \cdots, 0}_{\text{pq}}\ket{\mathbf{0}}_{\text{rq}}\\
    \rightarrow &\ket{q_1,\cdots, q_N}_{\text{sq}}\ket{q_j, q_j, \cdots, q_j}_{\text{pq}}\ket{\mathbf{0}}_{\text{rq}}.
\end{split}
\end{equation}

As shown in Figure~\ref{qc_copy_pa}, the parity of a single qubit can be copied to \( L \) qubits using only \( \mathcal{O}(\log(L)) \) layers of CNOT gates.  

\( U_{\text{LP}}(j) \) is the load-parity circuit illustrated in Figure~\ref{fig_pqc}. The \( U_{\text{LP}} \) operation can be expressed as  
\begin{equation}
    U_{\text{LP}}(j) = \prod_{l \in S(j)} CNOT(\text{pq}_l, \text{rq}_l),
\end{equation}
where $S(j) = \{l \mid Z_j \in P_l\}.$ We load the parity of the $j$-th system qubit onto the $l$-th rotation qubit only when $P_l$ contains $Z_j$.

\bibliography{GROP}

\end{document}